\documentstyle[prb,tighten,aps,epsf]{revtex}
\begin{document}
\draft
\title{Superconductivity in the Hubbard model with pair hopping}
\author{Stanis{\l}aw Robaszkiewicz~\cite{esr}}
\address{Department of Physics, A. Mickiewicz University,\\ul.
Umultowska 85, 61-614 Pozna\'n, Poland}
\date{Received \hspace{5mm} June 1996}
\date{\today}
\author { Bogdan R. Bu{\l}ka\cite{ebb}}
\address{Institute of Molecular Physics,
Polish Academy of Sciences,   \\
ul. Smoluchowskiego 17, 60-179 Pozna\'n, Poland}

\date{Received \hspace{5mm} }

\maketitle

\begin{abstract}

	The phase diagrams and superconducting properties
of the extended Hubbard model with pair hopping interaction, i.e. the
Penson-Kolb-Hubbard model are studied. The analysis of the model is performed
for $d$-dimensional hypercubic lattices, including $d=1$ and $d=\infty$, by
means of the (broken symmetry) Hartree-Fock approximations and, for
$d=\infty$, by the slave-boson mean-field method. For $d=1$, at half-filling
the phase diagram is shown to consist of nine different phases including
two superconducting states with center-of-mass momentum $q=0$ and $q=Q$
($\eta$-pairing), site and bond-located antiferromagnetic and charge-density wave
states as well as three mixed phases with coexisting
site and bond orderings. The stability range of the bond-type orderings is shrank
with increasing lattice dimensionality $d$ and for $d=\infty$ the corresponding
diagram consists of four phases only, involving exclusively site-located orderings.
Comparing the pair hopping model
with the attractive Hubbard model we found in the both cases gradual evolution from
the BCS-like
limit to the tightly bound pairs regime and a monotonic increase of the gap in
the excitation spectrum with increasing coupling. However, the dynamics of electron
pairs in both models is qualitatively different, which results in different
dependences of condensation energies and critical temperatures on interaction parameters
as well as in different
electrodynamic properties, especially in a strong coupling regime.
\end{abstract}

\section{INTRODUCTION}

	The purpose of the present work is the analysis of phase diagrams,
electronic orderings
and superconducting properties of the extended Hubbard model with pair hopping
interaction, i.e. the so-called Penson-Kolb-Hubbard (PKH) model,
\begin{eqnarray}\label{1}
H &=-t{\sum _{i,j,\sigma}}^{\prime}c_{i\sigma}^{\dag}c_{j\sigma}+
U\sum_in_{i\uparrow}n_{i\downarrow}\nonumber\\
&-J{\sum _{i,j}}^{\prime}c_{i\uparrow}^
{\dag}c_{i\downarrow}^{\dag}c_{j\downarrow}c_{j\uparrow}-{\mu}\sum_{i,\sigma}
n_{i\sigma},
\end{eqnarray}
where the prime over the sum means restriction to nearest neighbor (n.n) sites,
$t$ denotes the single electron hopping integral, $U$ is the onsite density-density
interaction, $J$ is the pair hopping (intersite charge exchange) interaction
and $\mu$ is the chemical potential. In the absence of the $U$ term the Hamiltonian
(\ref{1}) reduces to the Penson-Kolb (PK) model.~\cite{pk}

	We will treat the parameters $t$, $U$, $J$ as the effective (phenomenological)
ones, assuming that they include all the possible contributions and renormalizations
like those coming from the strong electron-phonon couplings or from the coupling
between electrons and other electronic subsystems in solid or chemical complexes~\cite{MRR}
(such that the values of $U$ and $J$ can be effectively either positive or negative).
It is notable that formally $J$ is one of the off-diagonal terms of the
Coulomb interaction
$-J= (ii|e^2/r|jj)$,~\cite{hub} describing a part of the so-called bond-charge
interaction, and the sign of the {\it Coulomb-driven} charge exchange is typically
negative(repulsive, $J<0$). However, the effective attractive interaction of
this form ($J>0$) is also possible~\cite{fra,rob87,bas} and in particular it
can originate from the coupling of electrons with intersite (intermolecular)
vibrations via modulation of the hopping integral,~\cite{fra} or from the on-site
hybridization term in a generalized periodic Anderson model.~\cite{rob87,bas}

	The PKH model is one of the conceptually
simplest phenomenological models for studying correlations and for description of superconductivity
of the narrow band systems with short-range, almost unretarded pairing. It
includes a nonlocal pairing mechanism (the pair hopping term $J$) that is distinct
from the on-site interaction in the attractive Hubbard model and that is the driving
force of pair formation and also of their condensation.  Thus, the
superconducting properties and the evolution from the Cooper pair regime to
the strong coupling local pair regime can be essentially different in these
two models.

	While most of the basic properties of the attractive Hubbard model
seems to be at present well understood after several years of intense studies,
the PKH model has been investigated only in a few particular
limits.~\cite{pk,aff,don,buzatu,sik,roy,bos,bo,roy97,jap}
The main
efforts concerned the ground state phase diagram of the half-filled
one dimensional PKH~\cite{don,buzatu,jap} and PK~\cite{aff,sik,bos,bo} models.
In the case of the PKH model these problems were studied by both, momentum-space
renormalization-group (MSRG) and the finite-size (exact diagonalization of finite-size
cells)
methods (for $U,J>0$),~\cite{don} by the real space
renormalization-group (RSRG) (for $U>0$),~\cite{roy}, by the continuum-limit
field theory (CFT) approach~\cite{jap} (for $U>0$) and within the Green's function
formalism in the mean-field approximation~\cite{buzatu}. However, in all these studies,
except [\onlinecite{jap}], the possibility for the bond-located orderings was not
considered and the exact form of the phase diagram in the whole range of
parameters $-\infty<U/t, J/t<\infty$ has not been established up to now.
The properties of the PKH model for higher dimensional lattices ($1<d\le\infty$)
and arbitrary electron concentration ($0<n<2$) have not been studied yet,
except for the limiting case of zero bandwidth.~\cite{srgp} The latter limit was
analyzed by the variational approach, in which the $U$
term is treated exactly and the intersite $J$ term - within mean-field approximation~\cite{srgp}
(such an approach yields exact results for $d=\infty$).

	In the paper we will study the PKH model for the case of $d$-dimensional
hypercubic lattices ($1\le d\le\infty$) and arbitrary, positive as well as
negative, $U$ and $J$. In the analysis we will apply a broken symmetry Hartree-Fock
approximation (HFA) (Sec.2) supplemented, for $d=\infty$, by the slave boson mean-field
approach (SBMFA) (Sec.3).
In the case of the Hubbard model and its various
extensions~\cite{MRR} the former approach is known to give credible results
at $T=0$ for any $U$ as far as the energy of the ground state and energy gap in
the ordered states is concerned. It usually provides qualitatively correct ground state phase
diagrams for arbitrary dimensions if all the proper broken
symmetry phases are included into the analysis. Moreover, for the electronic models
with intersite interactions only, the HFA becomes an exact theory in the limit
of infinite dimension ($d=\infty$). At $T>0$ the HFA is much less reliable, especially
for low dimensional systems and the limits of strong coupling, as it
neglects short-range correlations and the effects of collective excitations.
An obvious weakness of the HFA (both at $T=0$ and $T>0$) is inadequate description
of the normal (nonordered) phase. This failure is a consequence of the fact that the HFA
greatly overestimates the energy of the phases without long-range order. Going
beyond the HFA we will use the SBMFA. The slave-boson method is in principle
not restricted to weak or strong coupling and it is an improvement over the
former treatment since it takes into account local correlations.~\cite{rob}  We will
apply the SBMFA only for $d=\infty$, where the intersite coupling $J$ can be treated
adequately. For finite dimension ($d<\infty$) the SBMFA treatment of intersite
interactions is technically involved and to our knowledge it has not
been analyzed consistently so far.

\section{General formulation and the Hartree-Fock analysis}

	In the system considered several types of superconducting,
magnetic and charge orderings can develop. In the following we will study
the case of alternating (hypercubic)
lattices with nearest-neighbor single electron hopping $t$ and pair hopping $J$, and
restrict our considerations to the
one- and two-sublattice orderings,~\cite{com} described by the following order
parameters: the superconducting with the s-type (S) and the $\eta$-type ($\eta$)
pairing:
$x_{S}=\frac{1}{N}\sum_i\langle c_{i\downarrow}c_{i\uparrow}\rangle =
\frac{1}{N}\sum_k\langle c_{-k\downarrow}c_{k\uparrow}\rangle$,
$x_{\eta}=\frac{1}{N}\sum_ie^{i{\bf QR}_i}\langle c_{i\downarrow}c_{i\uparrow}\rangle =
\frac{1}{N}\sum_k\langle c_{-k+Q\downarrow}c_{k\uparrow}\rangle$;
the antiferromagnetic (AF) with the staggered magnetization located on sites
(sAF) or on bonds between sites (bAF):
$x_{sAF}=\frac{1}{2N}\sum_{i,\sigma}\sigma e^{i{\bf QR}_i}\langle c_{i\sigma}^
{\dag}c_{i\sigma}\rangle$,
$x_{bAF}=\frac{1}{2N}\sum^{\prime}_{i,j>i,\sigma}\sigma e^{i{\bf QR}_i}\langle c_{i\sigma}^{\dag}
c_{j\sigma}\rangle =
\frac{1}{4N}\sum_{k,\sigma}\sigma\eta_{k}\langle c_{k\sigma}^{\dag}c_{k+Q\sigma}\rangle $;
the charge density wave (CDW) with the on-site (sCDW) or the bond zigzag
(bCDW) modulation of charges:
$x_{sCDW}=\frac{1}{2N}\sum_{i,\sigma}e^{i{\bf QR}_i}\langle c_{i\sigma}^{\dag}c_{i\sigma}\rangle$,
$x_{bCDW}=\frac{1}{2N}\sum^{\prime}_{i,j>i,\sigma}e^{i{\bf QR}_i}\langle c_{i\sigma}^{\dag}c_{j\sigma}\rangle =
\frac{1}{4N}\sum_{k,\sigma}\eta_{k}\langle c_{k\sigma}^{\dag}c_{k+Q\sigma}\rangle $,
where $\eta_{k}=i\sum_a\sin(ka)$ and $Q=(\pi/a,\pi/a,...)$. We assume that the
sites are ordered in an ascending way along the crystallographic axis and for
the case of the bond zigzag parameters the sum is restricted to the nearest
neighbor sites j, which followed the i-th site. The number of electrons per lattice site is given by
$n=\frac{1}{N}\sum_{i,\sigma}\langle c_{i\sigma}^{\dag}c_{i\sigma}\rangle
$.
In the case of the AF phase we quoted above only sAF$_z$ and bAF$_z$ orderings,
corresponding to a $z$-component magnetization located on sites and bonds,
respectively, and we omitted s(b)AF$_x$, s(b)AF$_y$. Due to the
SU(2) - spin symmetry of the PKH model the latter orderings are strictly
degenerated with s(b)AF$_z$.

	Within the framework of the broken-symmetry Hartree-Fock approach
the mean-field Hamiltonian in the
momentum space $k$, including all types of orderings is given by
\begin{eqnarray}\label{4}
H_{HF} =\sum _{k,\sigma}(\epsilon_k-\mu
+\frac{U}{2}n)c_{k\sigma}^{\dag}c_{k\sigma}+
(U-zJ)\sum_k(x_S c_{k\uparrow}^{\dag}c_{-k\downarrow}^{\dag}+h.c.)\nonumber\\
 +(U+zJ)\sum_k(x_{\eta}c_{k\uparrow}^{\dag}c_{-k+Q\downarrow}^{\dag}+h.c.)-Ux_{sAF}
\sum_{k,\sigma} \sigma c_{k\sigma}^{\dag}c_{k+Q\sigma}\nonumber\\+
\frac{2J}{z}x_{bAF}\sum_{k,\sigma}\sigma\eta_kc_{k\sigma}^{\dag}c_{k+Q\sigma}
+Ux_{sCDW}\sum_{k,\sigma} c_{k\sigma}^{\dag}c_{k+Q\sigma}\nonumber\\-
\frac{2J}{z}x_{bCDW}\sum_{k,\sigma}\eta_kc_{k\sigma}^{\dag}c_{k+Q\sigma}\;,
\end{eqnarray}
where $\epsilon_k=-\tilde{t}\gamma_k$, $\tilde{t}=t+2pJ/z$, $z$ is the number
of nearest neighbor sites (for the hypercubic lattice of $d$-dimension: $z=2d$), and $p$
denotes the Fock term:
$p=\frac{1}{4N}\sum^{\prime}_{i,j,\sigma}\langle c_{i\sigma}^{\dag}c_{j\sigma}\rangle
= \frac{1}{4N}\sum_{k,\sigma} \gamma_k\langle c_{k\sigma}^{\dag}c_{k\sigma}\rangle $,
with $\gamma_k=\sum_a\cos(ka)$.

The eigensolutions of the Hamiltonian (\ref{4})
and the corresponding free energy
\begin{equation}\label{5}
F=-\frac{1}{\beta}\ln[Tr\{\exp(-\beta H_{HF})\}]+\langle H-H_{HF}\rangle _{_{HF}}+\mu N_e\;,
\end{equation}
where $\beta=1/k_BT$ and $N_e$ denotes the number of electrons in the system,
can be determined by the standard methods~\cite{HFA} with either the
Green's function or  the equation of motion approach.  If the solutions corresponding
to the pure phases (i.e. the phases with
only one type of order) are analyzed, the free energy (\ref{5}) may be expressed
in terms of the eigenvalues of $H_{HF}$ in the form
\begin{eqnarray}\label{5a}
\frac{F}{N}=\overline{\mu}(n-1)+\frac{U}{4}n^2+\frac{4}{z}Jp^2
+A_{\alpha}|x_{\alpha}|^2\nonumber\\-\frac{1}{\beta N}
\sum_{k,r}\ln[2\cosh(\frac{
\beta E^{r}_{\alpha\;k}}{2})]\;,
\end{eqnarray}
 where $r=\pm$, $\overline{\mu}=\mu-Un/2$, $V_{\alpha}$ is an effective coupling
strength for the $\alpha$ phase,
which is $V_S=-U+zJ$, $V_{\eta}=-U-zJ$, $V_{sAF}=U$, $V_{bAF}=-2J/z$,
$V_{sCDW}=-U$ and $V_{bCDW}=2J/z$ , $A_{\alpha}=V_{\alpha}$ for $\alpha=$
S, $\eta$, sAF, sCDW and $A_{\alpha}=2V_{\alpha}$ for $\alpha=$ bAF, bCDW.
The electronic spectrum is
$E^{\pm}_{S\;k}=\pm \sqrt{(\epsilon_k-\overline{\mu})^2+V_S^2x_S^2}$,
$E^{\pm}_{\eta\;k}=\epsilon_k\pm \sqrt{\overline{\mu}^2+V_{\eta}^2x_{\eta}^2}$,
$E^{\pm}_{sAF\;k}=\overline{\mu}\pm \sqrt{\epsilon_k^2+V_{sAF}^2x_{sAF}^2}$,
$E^{\pm}_{bAF\;k}=\overline{\mu}\pm \sqrt{\epsilon_k^2+V_{bAF}^2|\eta_kx_{bAF}|^2}$,
$E^{\pm}_{sCDW\;k}=\overline{\mu}\pm \sqrt{\epsilon_k^2+V_{sCDW}^2x_{sCDW}^2}$ and
$E^{\pm}_{bCDW\;k}=\overline{\mu}\pm \sqrt{\epsilon_k^2+V_{bCDW}^2|\eta_kx_{bCDW}|^2}$
for the S-, the $\eta$-, the sAF-, the bAF-, the sCDW- and the bCDW phases,
respectively. In the derivation of the eigensolutions we have assumed an alternated
lattice, i.e. $\epsilon_{k+Q}=-\epsilon_k$.

	For arbitrary electron concentration $n$ the stable solutions are
determined as the minimum of $F$ with respect to
the variational parameters $x_{\alpha}$ ($\alpha=$ S, $\eta$, sAF,
bAF, sCDW, bCDW), $p$ and $\mu$, i.e. by the equations
\begin{eqnarray}\label{6}
\partial F/\partial x_{\alpha}=0\;,\partial F/\partial p=0\;,\partial F/\partial \mu = 0\;.
\end{eqnarray}

	Besides the pure phases there are also solutions for various mixed type
orderings. We have analyzed the stability conditions for all such states and
found that some of them can be stable in a definite range of parameters. They
are summarized in the Table I together with the corresponding order parameters.
For example, we present here the equations describing the mixed s+bAF phase.
In this case the free energy (\ref{5}) is expressed in terms of the
eigenstates as
\begin{eqnarray}\label{12}
\frac{F}{N}=\overline{\mu}(n-1)+\frac{U}{4}n^2
+\frac{4}{z}Jp^2+ V_{sAF}x_{sAF}^2\nonumber\\+
2V_{bAF}|x_{bAF}|^2
-\frac{1}{\beta N}\sum_{k,r}\ln[2\cosh(\frac{\beta E^{r}_{sbAF\;k}}{2})]\;
\end{eqnarray}
with the electronic spectrum $E^{\pm}_{sbAF\;k}=\overline{\mu}\pm
\sqrt{\epsilon_k^2+ V_{sAF}^2x_{sAF}^2+ V_{bAF}^2|\eta_kx_{bAF}|^2}$ and
$x_{sAF}$, $x_{bAF}$, $p$ and $\mu$ are determined by a set of self-consistent
equations: $\partial F/\partial x_{sAF}=0$, $\partial F/\partial x_{bAF}=0$,
$\partial F/\partial p=0$ and $\partial F/\partial \mu = 0$.

\begin{figure}
\epsfxsize=13cm
\epsffile{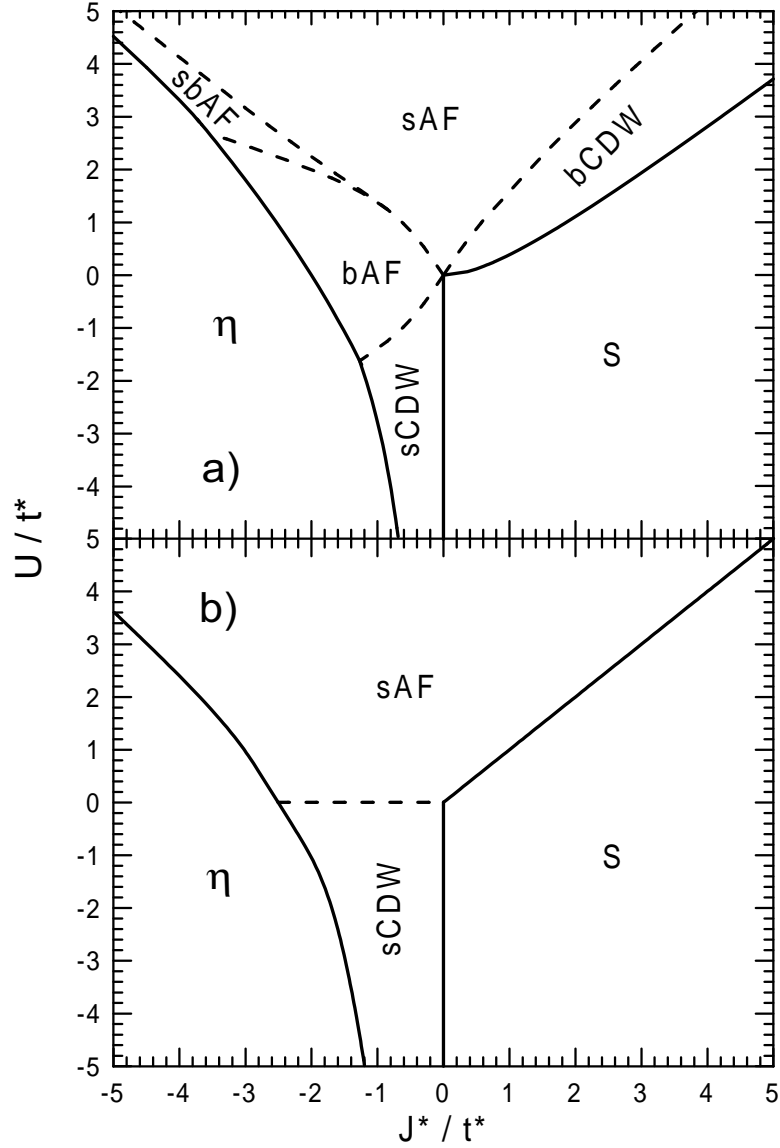 }
\caption{Phase diagram of the half-filled PKH model
for the 1D chain (Fig.a) and for the $d=\infty$-hypercubic lattice (Fig.b)
determined within the broken symmetry HFA.
The region of the mixed AF state is denoted by sbAF. Close to the boundary
lines separating the sAF and bCDW states as well as the sCDW and bAF
states there are very narrow regions (narrower than thickness of the curves in the
figure) of the stable mixed ordered phases (sAF+bCDW, for $U>0$ and sCDW+bAF for $U<0$).
First-order and second-order transition
phase boundaries are marked by solid and dashed curves, respectively. The SBMFA phase
diagram for $d=\infty$ is almost identical with Fig.b
(see discussion in Sec.3).
}
\label{f1}
\end{figure}

	 In order to determine the mutual stability of the
phases considered one has to find all the possible solutions and compare the
corresponding free energies. In the weak and strong coupling regimes we were
able to derive several analytical expressions concerning the energy gaps, the
order parameters and the critical temperatures, but in a general
case numerical methods had
to be used. At $T=0$ we performed complete numerical analysis
of all the solutions for the whole range of the parameter values and the resulting
phase diagrams are presented in Fig.\ref{f1} for the 1D chain and the hypercubic
lattice of the dimension $d=\infty$. The renormalized parameters are: $J^*=Jd$ and
$t^*=t\sqrt{d}$.

	For $d=\infty$ the density of states (DOS) is
$\rho(\epsilon)=\exp[-\epsilon^2/(8{t^*}^2)]/(\sqrt{8\pi}t^*)$.
In this case there are not stable states with the bond
type of ordering as all bond parameters disappear in the limit $d\to \infty$.
Also, the Fock term $p$ is then irrelevant as the effective width
of the electronic band $W_{eff}\equiv 4\tilde{t}d = 4t^*\sqrt{d}+4J^*p/d$
and the second term disappears for $d=\infty$. This is in contrast to the $d=1$
case (Fig.\ref{f1}a), where the AF and CDW orderings of the bond type can
exist in a wide range of parameters (the former for $J<0$ and the latter for $J>0$).
The bond type ordering can also coexist with the on-site type ordering, as
it is seen in Fig.\ref{f1}a for the mixed s+bAF phase.
There have been also found very narrow regions of the stable mixed phases:
bAF+sCDW (for $U<0$) and sAF+bCDW (for $U>0$). The
curves separating the sAF- and the bAF-type orderings are the lines of
second order phase transition, at which the parameter $x_{sAF}$ or
$x_{bAF}$ disappears. In the lattices of dimension $1<d<\infty$ one can analyze
more complex bond orderings (e.g. the phase of fluxes), however, the ranges of
stability of all the bond-ordered phases will be gradually shrank with increasing
lattice dimension.

\begin{figure}
\epsfxsize=13cm
\epsffile{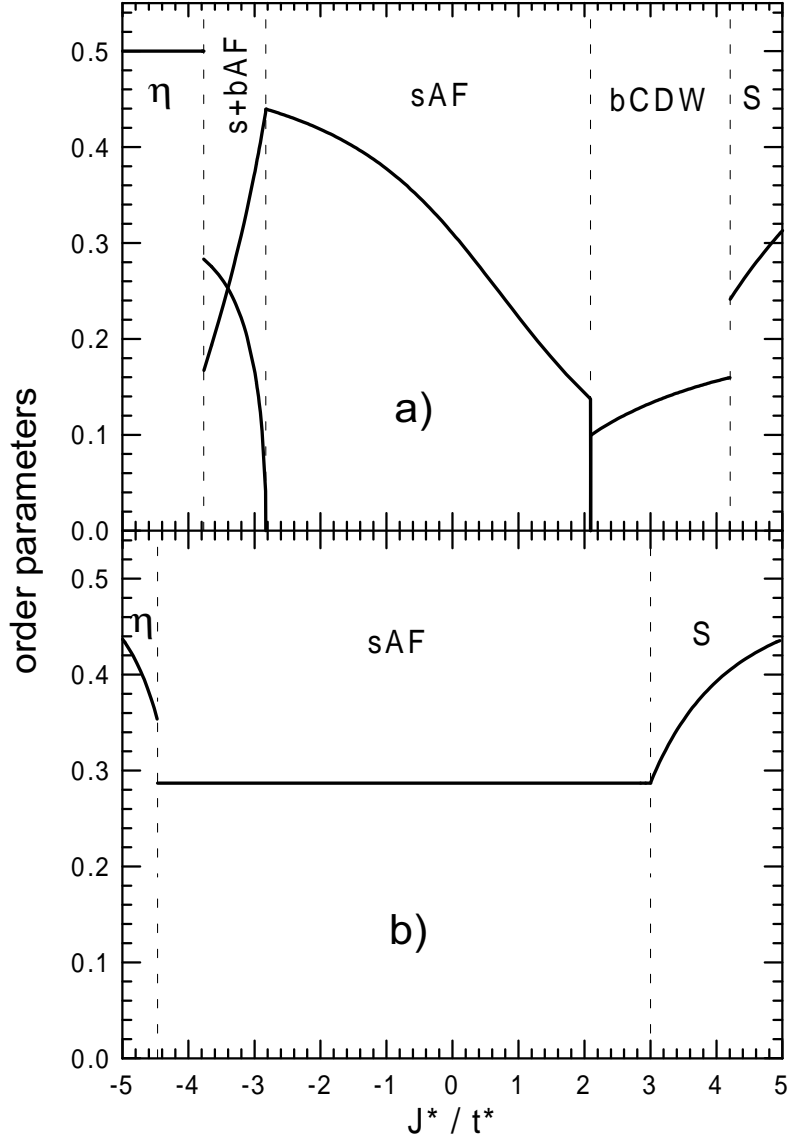 }
\caption{The $J^*/t^*$ dependence of the order parameters in the ground state of
the $d=1$ (Fig.a) and $d=\infty$ (Fig.b) system, for $n=1$ and $U/t^*=3$. The
stability ranges of the different phases are indicated by the vertical dashed lines.
The mixed sAF+bCDW state (Fig.a) exists for $ J^*/t^*\in (2.09105, 2.09425)$.}
\label{f2}
\end{figure}

	The $J^*$ dependence of the order
parameters for $U/t^*=3$ is presented in Fig.\ref{f2}, where the upper part
is for the $d=1$ system and the lower part for $d=\infty$.
Fig.\ref{f2}a shows a wide
range of the mixed s+bAF phase with $x_{sAF}\ne 0$ and $x_{bAF}\ne 0$.
The parameter $x_{bAF}\to 0$ for $J^*/t^*\to -2.83$, indicating the
second-order transition.  In the case presented in Fig.\ref{f2}a ($U/t^*=3$)
the mixed sAF+bCDW phase is stable only in a very
narrow range $2.09105<J^*/t^*<2.09425$.

\section{SLAVE-BOSON STUDIES}

	In the previous paper~\cite{rob} we showed that the slave boson
mean-field approach (SBMFA) gives reliable results for the ground state
properties of the attractive Hubbard model in the whole range of coupling $|U|$
and arbitrary electron concentration $n$. Therefore, we also
applied this method to the present model (\ref{1}). As the SBMFA takes into
account the onsite electron correlations and neglects
the short-range intersite correlations (the Fock term and the bond
type orderings are omitted), we have concentrated on the case
of $d=\infty$ lattice, where the mean field treatment of intersite interactions
becomes exact.

	In the slave-boson approach each local state is described by a fermi
operator $f_{i\sigma}$ and two types of bose operators $\bbox{p}_i$ and
$\bbox{b}_i$, which correspond to two vector fields: a field of local magnetic
moments and that of local charges. The completeness condition means that length
and direction of the vectors  $\bbox{p}_i$ and $\bbox{b}_i$ can vary from site
to site, but a sum of their length is always $p_i^2+b_i^2=1$. We use the spin-
and the charge-rotationally
invariant slave-boson representation~\cite{FW,rob}, in which the order
parameters  are expressed by
$x_{S}=\frac{1}{N}\sum_{i,\rho}\langle b_{ix}^{\dag}b_{ix}+b_{iy}^{\dag}b_{iy}\rangle$,
$x_{\eta}=\frac{1}{N}\sum_{i,\rho}e^{i{\bf QR}_i}\langle b_{ix}^{\dag}b_{ix}+b_{iy}^{\dag}b_{iy}\rangle$,
$x_{sCDW}=\frac{1}{N}\sum_{i} e^{i{\bf QR}_i}\langle b_{iz}^{\dag}b_{iz}\rangle$,
and $x_{sAF}=\frac{1}{N}\sum_{i} e^{i{\bf QR}_i}\langle p_{iz}^{\dag}
p_{iz}\rangle$, for the superconducting, sCDW and sAF phase, respectively.
In the mean field studies we confine ourselves to the temperature
$T=0$ and neglect  space and time fluctuations of the bose fields. The operators $\bbox{p}_i$ and
$\bbox{b}_i$ are replaced by their expectation values, which in the following
are treated as variational parameters. The SBMFA is, therefore, a variational
method on a trial state described by the Hartree-Fock wave functions (being
equivalent to the Gutzwiller approximation).~\cite{KR} The free energy is
the sum of the fermionic and bosonic parts, and for the $\alpha$ phase,
where $\alpha=$ S, $\eta$, sCDW, sAF,  and any given $n$ it can be written in the
following unified form:
\begin{eqnarray}\label{13s}
\frac{F^{SBMFA} }{N}= \frac{F_f}{N} +\frac{F_b}{N}= \nonumber\\
 -\frac{1}{N\beta}\sum_{k,r} [\ln\lbrace 1+ \exp(\beta E^{sb\;r}_{\alpha\;k})\rbrace]
+ \frac{U}{2}(b^2+2\delta)\nonumber\\
+C_{\alpha}-(\lambda_0+\mu)(1+2\delta)-2\lambda_{\alpha}x_{\alpha}\;,
\end{eqnarray}
where $r=\pm$, $C_S=-2J^*x_S^2$, $C_{\eta}=2J^*x_{\eta}^2$, $C_{sCDW}=0$, $C_{sAF}=0$,
$b^2 = \langle b_{ix}^{\dag}b_{ix}+b_{iy}^{\dag}b_{iy}+
b_{iz}^{\dag}b_{iz}\rangle$ and
$2\delta = n-1 =\langle b_{iz}^{\dag}b_{iz}\rangle$,
$\lambda_0$ and $\lambda_{\alpha}$ are the Lagrange multipliers. The fermionic
spectrum is $E^{sb\;\pm}_{S\;k}= \pm\sqrt{(q_S\epsilon_k+\lambda_0)^2+\lambda_S}$,
$E^{sb\;\pm}_{\eta\;k}= q_{\eta}\epsilon_k \pm\sqrt{\lambda_0^2+\lambda_{\eta}^2}$,
 $E^{sb\;\pm}_{sCDW\;k}=-\lambda_0 \pm\sqrt{q_{sCDW}\epsilon_k^2+\lambda_{sCDW}^2}$
and $E^{sb\;\pm}_{sAF\;k}=-\lambda_0 \pm\sqrt{q_{sAF}\epsilon_k^2+\lambda_{sAF}^2}$.
Its $k$-dependence is analogous to that obtained in the HFA with
the bandwidth reduced by the factor
\begin{equation}\label{15s}
q_{\alpha} = \frac{2p^2(b^2+\sqrt{b^4-4x_{\alpha}^2-4\delta^2})}{1-4x_{\alpha}^2-4\delta^2}\;,\\
\end{equation}
for $\alpha=$ S, $\eta$, sCDW, and
\begin{equation}\label{15bs}
q_{sAF}= \frac{2b^2(p^2+\sqrt{p^4-4x_{sAF}^2-4\delta^2})}{1-4x_{sAF}^2-4\delta^2}\;.\\
\end{equation}
The stable solutions are determined from the minimum of the free energy
$F^{SBMFA}$ with respect to $x_{\alpha}$, $\lambda_{\alpha}$,
$\lambda_0$ and $b$.

	 In determination of the phase diagram of the half-filled PKH model we
compare the free energies $F^{SBMFA}_{\alpha}$ corresponding to the S, $\eta$, sCDW and sAF phases.
Their values are different from those obtained in the HFA and depend on the band
narrowing factors $q_{\alpha}$. These factors are important parameters. In the
normal phase
($x_{\alpha}=0$) the band narrowing process can lead to insulating phase
($q_N=0$) for large coupling $|U|\gg t$.~\cite{rob,KR,FW,bak} However, for
alternating lattices and $n=1$ the normal phase
is not a ground state as its free energy is always higher than that of the long-range
ordered phases (S, $\eta$, sCDW and sAF).  For all these phases the band narrowing
factors $q_{\alpha}$  are close to unity, for example, in the attractive Hubbard
model: $0.954<q_S <1$ in $d=\infty$ ( see also Ref.[\onlinecite{bak}]).
Thus, the SBMFA free energies of the ordered phases are relatively close to the corresponding
HFA results. The SBMFA phase diagram of the PKH model for
$d=\infty$ is, therefore, very similar to that given in Fig.\ref{f1}b. In
particular, the location of the S-sAF phase boundary in the ground
state phase diagram can be expressed most conveniently in terms of the deviation
$\epsilon_c=J^*/U-1$ from the line $J^*/U=1$.  Within the HFA, the S-sAF phase boundary
is given by $J^*/U=1$ for any $t^*$, and for $d=\infty$ it agrees with a rigorous
solution at $t^*=0$.~\cite{srgp} Within the SBMFA, $\epsilon_c$ is found to depend
sensitively on the strength of the interactions and one obtains that $\epsilon_c>0$ for any
$\infty>t^*>0$, with a maximum deviation $\epsilon_c\approx 0.02$ found for
$U/t^*\approx 3.5$ and with $\epsilon_c\to 0$ for $t^*\to 0$ as well as
$t^*\to \infty$. It means that the hopping term slightly extends the stability
range of the sAF phase with respect to the S phase. Notice that similar results
are obtained for the extended Hubbard model with nearest neighbor density-density
repulsion $W^*$. In that case Monte-Carlo simulations~\cite{hir} and perturbational
treatments~\cite{vandong} show that for $t\ne 0$ the actual phase boundary is also
slightly shifted upward relative to the line $W^*/U=1$ predicted by the HFA.

\begin{figure}
\epsfxsize=11cm
\epsffile{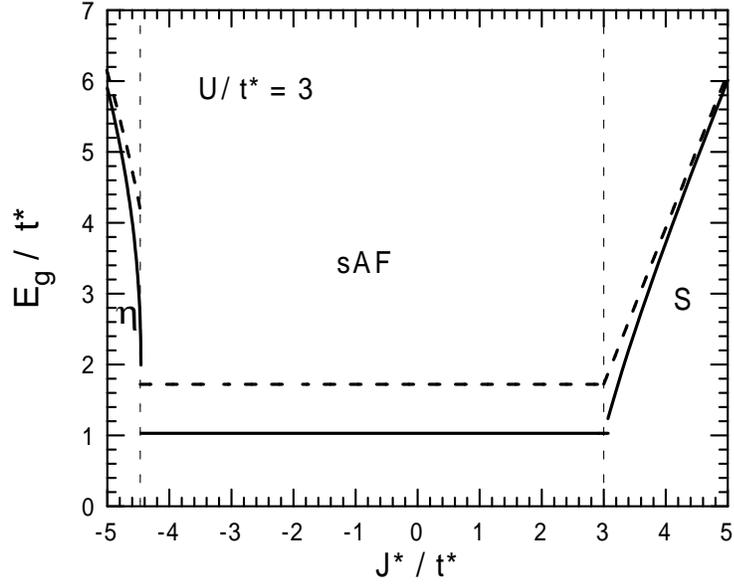 }
\caption{The dependence of the gap in the excitation spectrum on $J^*/t^*$
calculated in the SBMFA (solid curves) and in the HFA (dashed
curves) for $d=\infty$ hypercubic lattice, $U/t^*=3$ and  $n=1$. The
stability ranges of the different phases are indicated by the vertical dashed lines.}
\label{f3}
\end{figure}

\begin{figure}
\epsfxsize=13cm
\epsffile{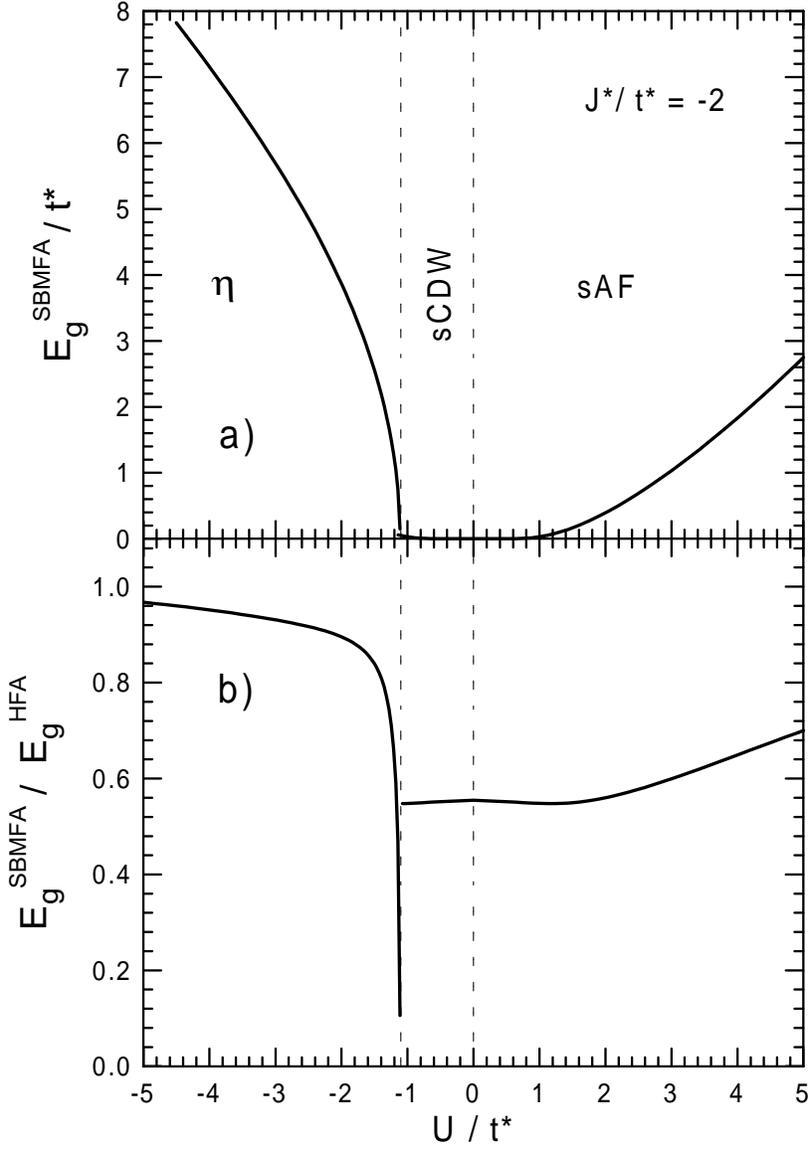 }
\caption{The plots of the gap in the excitation spectrum (Fig.a) and the
reduction factor $\gamma_{\alpha}=E_g^{SBMFA}/E_g^{HFA}$ (Fig.b) as
a function of $U/t^*$, calculated in the SBMFA for
$d=\infty$ hypercubic lattice, $J^*/t^*= -2$ and $n=1$.}
\label{f4}
\end{figure}

	Although the SBMFA gives minor changes in the ground state energies,
other physical characteristics are modified in a much more pronounced way.
We will show it analyzing the gap $E_g$ in the excitation spectrum determined within
the SBMFA  as well as the HFA. Fig.\ref{f3} shows dependences
of $E_g$ on $J^*$ in the case of $U/t^*=3$ and $n=1$. The value of
$E_g^{SBMFA}$ is reduced with respect
to $E_g^{HFA}$. The results are closer to each other for larger couplings
$|J^*|$, where the onsite correlations become less relevant. The maximum reduction
is seen for  the gap in the $\eta$ state, which at the transition line is reduced
by a factor $\gamma_{\eta}\equiv E_{g\;\eta}^{SBMFA}/E_{g\;\eta}^{HFA}=0.56$.
Fig.\ref{f4} presents the $U/t^*$ dependence of $E_g$  for
$J^*/t^*=-2$ and $n=1$. The gaps $E_{g\;sAF}$ and $E_{g\;sCDW}$
corresponding to the sAF and sCDW phases do not depend on $J^*$, and
they are the same as in the usual Hubbard model ($J^*=0$).  In the lower part of
Fig.\ref{f4} the reduction parameter $\gamma_{\alpha}$ is shown. The minimum value of
$\gamma_{\eta}$ is $0.11$ for $E_{g\;\eta}^{SBMFA}$ close to the transition
point to the sCDW phase. The energy gaps $E_{g\;sAF}^{SBMFA}$ and $E_{g\;sCDW}^{SBMFA}$ are
maximally reduced for a weak coupling $|U/t^*|\ll 1$.  In this limit one can find that
\begin{equation}\label{16s}
E_{g\;\alpha}^{HFA}= At^*\exp [-\frac{\sqrt{8\pi}t^*}{|U|}]
\end{equation}
and
\begin{equation}\label{17s}
\gamma_{\alpha}\equiv \frac{E_{g\;\alpha}^{SBMFA}}{E_{g\;\alpha}^{HFA}}
=\exp [-\frac{3\pi}{16}]=0.554855\;,
\end{equation}
where $\alpha=$ sAF, sCDW and S,  $A=4\sqrt{2}e^{-\gamma/2}=4.23871$ and
$\gamma=0.577216$ is Euler gamma constant.
(The value (\ref{17s}) is larger than
$\gamma_S=\exp[-3/4]=0.472267$ obtained for the rectangular density of
states.~\cite{rob} )

\section{Discussion and Concluding remarks}

	Let us compare the properties of superconducting
phases and their evolution with a change of coupling and concentration for the
three limiting cases of the model (\ref{1}): i) the attractive Hubbard model with
$U<0, J=0$ , ii) the PK model with $U=0, J>0$, and iii) the PK model with
$U=0, J<0$. We will discuss qualitative differences and similarities in the
behavior of the system for these limits and stress distinct features of each
case.

	In the first two cases the pairing interaction favors the on-site s-wave
superconductivity (S), whereas in the third one - the $\eta$-pairing. Moreover,
the later two cases include a nonlocal pairing mechanism ($J$)
that is distinct from the zero-range instantaneous interaction existing in
the i) case. The difference between i) and ii) occurs in the case of the
half-filled band.  At $n=1$ the $U<0$ Hubbard model posses SU(2)
symmetry of the charge sector and
is characterized by coexistence of the sCDW and the S ordering (these
phases are strictly degenerated) in the ground state. No such degeneracy
occurs in the PK model, as its charge sector is governed by the $U(1)$
symmetry for any $n$.

\begin{figure}
\epsfxsize=11cm
\epsffile{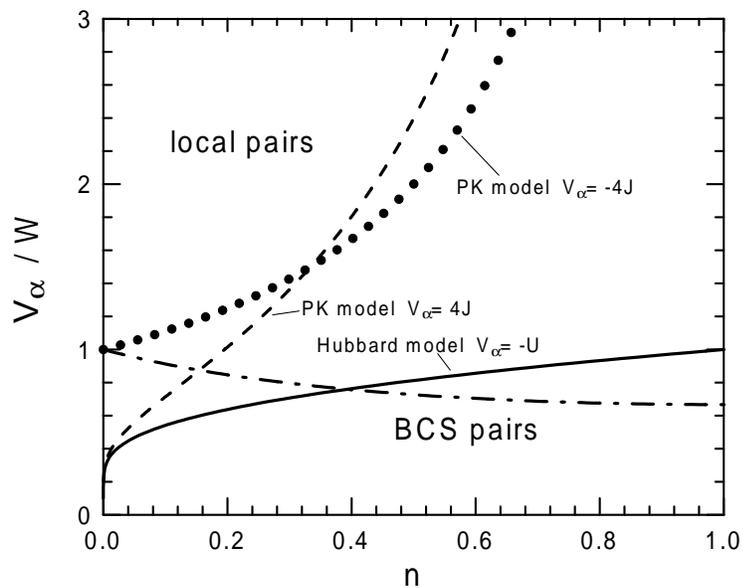 }
\caption{Boundary lines between the regions of the BCS-like and the local pair
superconductivity calculated as a function of $n$ in the case of $d=2$ lattice
for the $U<0$ Hubbard model
($V_{\alpha}=-U$) - solid curve, as well as for the PK model with $J>0$
($V_{\alpha}=4J$ ) - dashed curve, and with $J<0$ ($V_{\alpha}=-4J$) - dotted
curve. For the PK model with $J<0$ the $\eta$-type Cooper pairs are stable only above the long-short dashed.}
\label{f9}
\end{figure}

	For i) and ii) at $T=0$ the S-phase is stable for any nonzero
interaction ($U<0$ or $J>0$) and arbitrary $n$ ($0<n<2$). In both these cases the
evolution of the S-phase from the BCS like superconductivity, with extended
Cooper pairs, to superconductivity of composite bosons (local pairs)
with increasing coupling is continuous. At $T=0$ the appropriate boundary between
both regimes can be located (after Leggett [\onlinecite{leg}]) from the requirement
that the chemical potential in the superconducting phase reaches the bottom of the electronic
band, i.e. from $\mu_S=-W/2$. For
$d=2$ lattice the borderlines as a function of $n$ are shown in Fig.\ref{f9}.
As we see for both models with increasing $n$ the boundaries are shifted
towards higher values of coupling. For $d=3$ the corresponding plot has qualitatively
similar form, except $n\to 0$ limit, where there is a
critical value of coupling for pair formation.

	For the case iii) the $\eta$-phase is stable
only below a critical value of $J$ and for $d<\infty$ the local pair regime is
reached directly after crossing the $\eta$ phase boundary. The critical value
$J_c$ depends on the lattice structure, the lattice dimensionality
($d$) and the band filling ($n$). The estimations of $J_c$ for various cases
are collected in Table II. Except $d=\infty$, the transition at $J_c$ is of
the first order and characterized by an abrupt change in the structure
of the ground state. For $d=\infty$ the phase stable for $J_c<J<0$ is a normal
metal without any long-range ordering (for any $n$), whereas for $d<\infty$ and
$n=1$ that phase is insulating and antiferromagnetic with bond-type modulation
of magnetization.

\begin{figure}
\epsfxsize=13cm
\epsffile{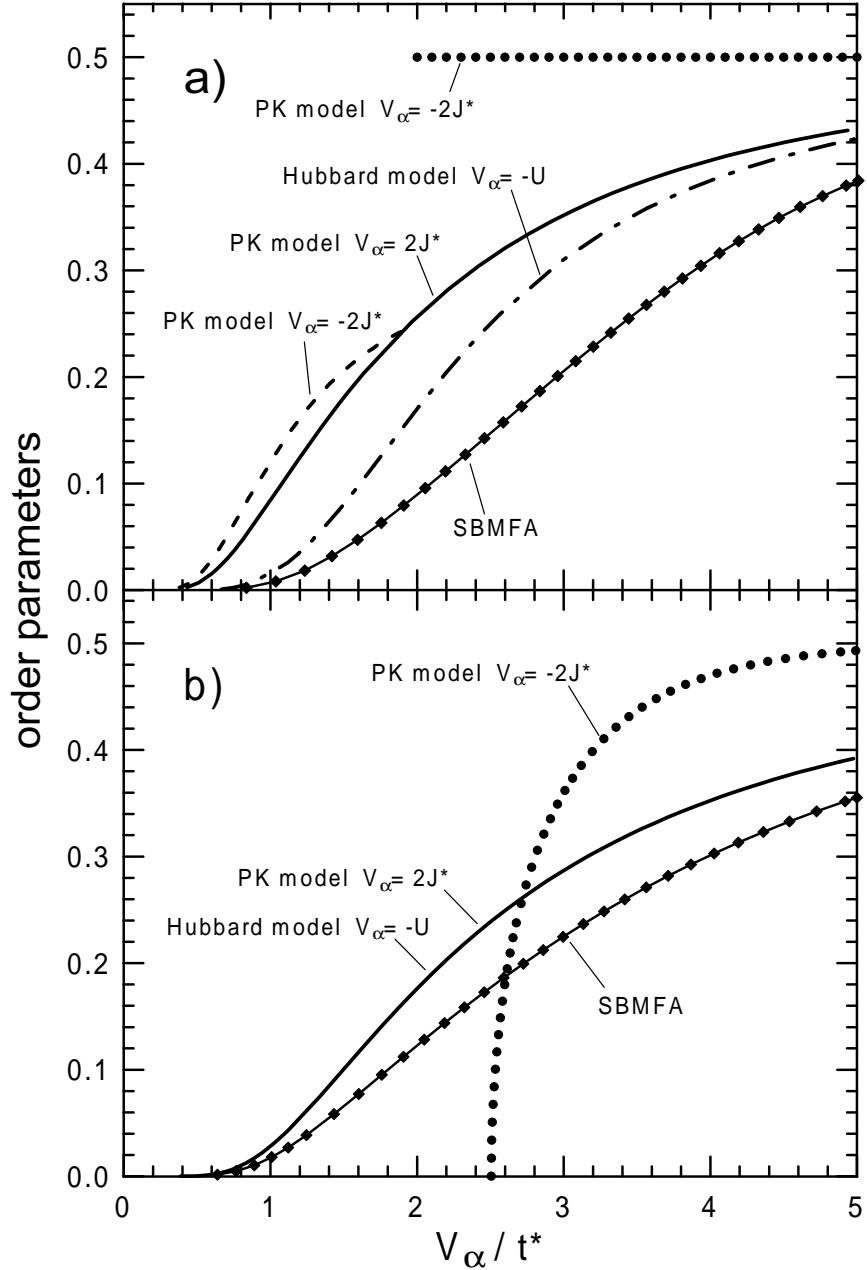 }
\caption{The plots of the gap parameters $x_{\alpha}= E_g^{\alpha}/V_{\alpha}$
as a function of
the coupling parameters $V_{\alpha}$ for the S phase of the attractive Hubbard
model (dashed-dot curve, $V_{\alpha}=-U$) and the PK model (solid curve,
$V_{\alpha}=2J^*$) as well as for the $\eta$ phase (dotted curve) and the bAF
phase (dashed curve) of the PK model ($V_{\alpha}=-2J^*$), calculated
 within the HFA, in the case of $d=1$ (Fig.a) and $d=\infty$ (Fig.b) lattices, $n=1$.
For $d=\infty$
the plots of $x_S$ vs. $V_{\alpha}$ for the PK model and the $U<0$ model have
the same form (solid curve). For the sake of comparison the SBMFA results for the $U<0$
Hubbard model in $d=1$ and $d=\infty$ are presented by the curves with diamonds.}
\label{f10}
\end{figure}

	The evolution of the gap parameters $x_{\alpha}$ ($\alpha=$ S, $\eta$
and bAF) with increasing interaction $V_{\alpha}$ ({\it for all three cases}) is
presented in Fig.\ref{f10} for $d=1$ (Fig.\ref{f10}a) and $d=\infty$
(Fig.\ref{f10}b). The corresponding
plots for $d=2$ and $d=3$ lattices have qualitatively the same form to
those for $d=1$. For the sake of comparison we
have shown also the SBMFA results (curves with diamonds in Fig.\ref{f10})
calculated for the attractive Hubbard model in $d=1$ and $d=\infty$.
Notice the first order transition from the bAF state to the
$\eta$ state in the $J<0$ PK model for $d=1$. In this case $x_{\eta}$ has
the maximum value 1/2. On the contrary for $d=\infty$ the phase transition to
the $\eta$-state is of a second-order, and $x_{\eta}$ continuously increases with
decreasing $J^*$ (it never saturates for a finite $J^*$). For the $J>0$ PK
model and the $U<0$ Hubbard model one observes a continuous evolution of the
order parameter $x_S$ with increasing $V_{\alpha}$ and an exponential
(BCS-like) behavior of $x_S$ in the weak coupling limit.

\begin{figure}
\epsfxsize=11cm
\epsffile{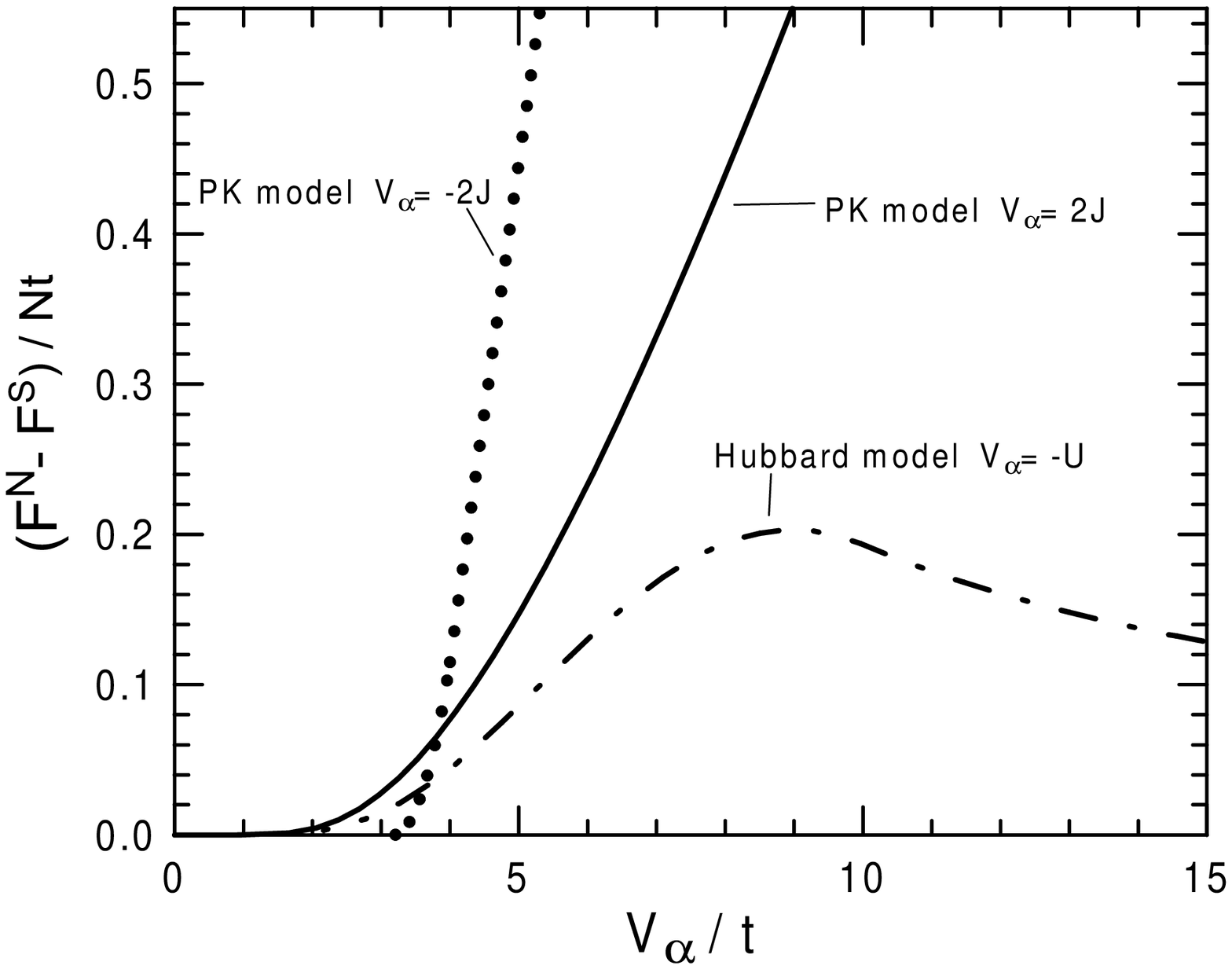 }
\caption{Difference between the free energies for the normal and the
superconducting phase plotted as a function of coupling parameters
$V_{\alpha}$ for the attractive Hubbard model (dashed-dot curve, $V_{\alpha}=-U$)
and the PK model (solid curve,
$V_{\alpha}=2J$) as well as the difference of the free energies for the normal and
the $\eta$-type superconducting phase for the PK model vs.  $V_{\alpha}=-2J$
(dotted curve). The derivations were performed for the $d=1$ chain at $n=0.8$
and $T=0$ by means of the SBMFA for the attractive Hubbard model and by the
HFA for the PK model.}
\label{f11}
\end{figure}

	As we have already pointed out, for both models there is a crossover
from the BCS-like limit to the tightly bound pairs regime with increasing
coupling and this evolution of the superconducting (S) phase is gradual.
However, the thermodynamic
and electromagnetic properties of both the models are very different beyond the
weak coupling limit.~\cite{rob,txujst} To illustrate the situation we have plotted
in Fig.\ref{f11} the condensation energies, i.e. the difference of the free energy  in the normal
and in the superconducting phase, $\Delta F= F^N-F^S$, as a function of the coupling
parameters $V_{\alpha}$. These results have been obtained for the $d=1$ chain at
$n=0.8$, but for the lattices of other dimension and other electron concentrations
one gets qualitatively similar dependences.
They are in good qualitative agreement with results of perturbational expansions
for the models considered both in the weak coupling
($V_{\alpha}/t\ll 1$) as well as   in
the strong coupling regimes ($V_{\alpha}/t\gg 1$).~\cite{rob,txujst}
 As the
square of the thermodynamic critical field $H_c^2$ is proportional to $\Delta F$,
we conclude that in the attractive Hubbard model this
quantity (similarly as the critical temperature $T_c$) increases exponentially
for small values of $|U|$, then
it goes through a round maximum and it decreases as $t^2/|U|$ for large
$|U|$. On the contrary, in the PK model we found no maximum of $H_c^2$
and $T_c$ at intermediate coupling and both these quantities increase linearly
with $J$ for large $J$. Also the behavior of the penetration depth
$\lambda_L$ and the pair mobility $t_p$ is
different. In the strong coupling limit $\lambda_L^2\propto 1/t_p$ {\it increases}
with $|U|$ in the former model ($\lambda_L^2\propto |U|/t^2$), while it
{\it decreases} with $J$ in the PK model ($\lambda_L^2\propto 1/J$).  It is in
agreement with studies of collective
excitations performed using a generalized random-phase approximation.~\cite{roy97}
The collective-mode velocity increases  with $J$ in the PK model, in contrast
to the attractive Hubbard model where it decreases with the coupling $|U|$.

	The phase diagram of the half-filled one-dimensional PK model ($U=0$,
$J\ne 0$) derived within the HFA is in agreement with that obtained by the density
matrix renormalization group
method.~\cite{aff,sik,bos} For $J>0$ both approaches  predict a continuous
second-order transition to usual s-wave pairing state at $J=0^+$, with no additional
transition for any $J>0$ (in contrast to the earlier predictions~\cite{don,roy}).
We have found that (at least for alternating lattices), this phenomenon remains
unchanged in higher dimensions (including the exactly solvable case of $d=\infty$)
and does not depend on the band filling. With increasing coupling there is a
gradual crossover from the BCS-like superconductivity to the superfluidity of
tightly bound local pairs. On the contrary, for $J<0$ the HFA predicts that
the $\eta$-phase is stable only above a critical value of $|J|$ and that the transition
at $J_c$ is of the first order for any $d<\infty$. Let us stress that for $n=1$
the values of $J_c$ calculated within  the HFA are in very good quantitative
agreement with the results of other more elaborated treatments available for
$d=1$ chain~\cite{sik,roy,bo}(cf. Table II). Moreover for $n\to 0$ the HFA yields exact
results for $J_c$ for any dimension.

	We have found that
the interplay between the on-site Coulomb interaction $U$ and the
intersite pair hopping $J$ in the PKH model can stabilize several new ordered
phases absent in
the usual Hubbard model ($U\ne 0$, $J=0$) and in the usual PK model ($U=0$, $J\ne 0$):
the bCDW and the mixed bCDW+sAF phases (for $U>0$, $J>0$), the mixed s+bAF phase (for
$U>0$, $J<0$) as well as the sCDW and the mixed bAF+sCDW phases (for $U<0$, $J<0$).
The new phases predicted by our broken symmetry HFA approach (which can be truly
long-range in $d\ge 2$) indeed need further
examination by more rigorous methods as the exact diagonalization of small
systems, density renormalization group, etc. We should point out however that
our findings concerning the bond-ordered solutions are clearly supported by
recent work of Japaridze and M{\"u}ller-Hartmann~\cite{jap} performed for the
$d=1$ PKH model with $U\ge 0$ in the weak coupling, using the continuum limit
field theory approach and bosonisation technique.  In contrast to previous
studies,~\cite{don,buzatu,roy} which have not considered the possibility
of bond-located orderings, present results and  those of Ref.[\onlinecite{jap}]
indicate that the bCDW (bAF) state but not the sAF or sCDW phases, is unstable
with respect to transition into the S ($\eta$) phase with increasing $J$ ($-J$)
for $U>0$.

	We have also compared the superconducting properties of the PK model
with those of the attractive Hubbard model. Although the energy gaps have
similar dependences on
the coupling parameters in the both models (see also [\onlinecite{bos}]),
dynamics of electron pairs is qualitatively different, which results in
different electrodynamic properties and different coupling dependences of $T_c$,
especially in a strong coupling regime.

\acknowledgments
	The paper is supported from the State Committee for Scientific Research
Republic of Poland within Grant No.~2 P03B 104 11 (S.R.) and 2 P03B 075 14
(B.R.B). We wish to thank R. Micnas for useful comments and discussions.

\newpage
\begin{table}
{TABLE I.} Phases considered and the corresponding order parameters.
\begin{tabular}{c|c}
Type of phase&Order parameters\\
\hline
S&$x_S\ne 0$\\
$\eta$&$x_{\eta}\ne 0$\\
sAF&$x_{sAF}\ne 0$\\
bAF&$x_{bAF}\ne 0$\\
s+bAF&$x_{sAF}\ne 0$, $x_{bAF}\ne 0$\\
sCDW&$x_{sCDW}\ne 0$\\
bCDW&$x_{bCDW}\ne 0$\\
s+bCDW&$x_{sCDW}\ne 0$, $x_{bCDW}\ne 0$\\
bAF+sCDW&$x_{bAF}\ne 0$, $x_{sCDW}\ne 0$\\
sAF+bCDW&$x_{sAF}\ne 0$, $x_{bCDW}\ne 0$\\
S+sCDW&$x_{S}\ne 0$, $x_{sCDW}\ne 0$\\
S+bCDW&$x_{S}\ne 0$, $x_{bCDW}\ne 0$\\
$\eta$+sCDW&$x_{\eta}\ne 0$, $x_{sCDW}\ne 0$\\
$\eta$+bAF&$x_{\eta}\ne 0$, $x_{bAF}\ne 0$\\
$\eta$+sAF+bAF&$x_{\eta}\ne 0$, $x_{sAF}\ne 0$, $x_{bAF}\ne 0$\\
\end{tabular}
\end{table}

\begin{table}
{TABLE II.} The HFA estimates of the critical value of $J$ below which
the $\eta$-state has lower energy than the normal state in the PK model with $J<0$.
In the limit $n\to 0$ the exact solution is $J_c^*=-W=-2\sqrt{d}t^*$. The results obtained
by density matrix renormalization group~\cite{sik} (DMRG), Lanczos diagonalization~\cite{bo}
 and real space renormalization-group~\cite{roy} (RSRG)  methods for the 1D chain are
given as well.
\begin{tabular}{l|c|c}
system	&\multicolumn{2}{c} {$J^*_c/t^*$}\\ \cline{2-3}
		&$n=1$	&$n\to 0$ (exact)\\ \hline\hline
$d=1$	&$-\pi/2\approx -1.5708$&$-2$ \\
		&$-1.5\;$~[\onlinecite{sik}] ,
$-1.75\;$~[\onlinecite{bo}], $-1.65\;$~[\onlinecite{roy}] &\\
\hline
$d=2\quad$ square lattice& 0& $-2\sqrt{2}\approx -2.8284$\\
$\quad\quad$rectangular DOS&$-8/3\sqrt{2}\approx -1.8856$&\\
\hline
$d=3\quad$ sc lattice&$-1.7028$&$-2\sqrt{3}\approx -3.4641$\\
$\quad\quad$elliptic DOS&$-3\sqrt{3}\pi/8\approx -2.0405$&\\
\hline
$d=\infty$&$-\sqrt{2\pi}\approx -2.5066$&$-\infty$
\end{tabular}
\end{table}

\begin{thebibliography}{99}
\bibitem{com} Obviously, in more general cases (i.e. for nonalternating lattices or
longer-ranged hoppings $t_{ij}$ as well as $n\ne 1$) the number of possible ordered
states can be much larger, including
also various types of incommensurate phases, phase of vortices, phase-separated
states, etc. We postpone discussion of this subject to the forthcoming paper.
\end{thebibliography}
\end{document}